\documentstyle[prl,aps,epsfig,floats,twocolumn]{revtex}
\begin{document}
\draft
\wideabs {
\title{Nonnormality and the localized control of extended systems}

\author{Roman O. Grigoriev, Andreas Handel}

\address{School of Physics, Georgia Institute of Technology,
         Atlanta, GA 30332-0430}

\date{\today} \maketitle

\begin{abstract}

The idea of controlling the dynamics of spatially extended systems
using a small number of localized perturbations is very appealing - such a
setup is easy to implement in practice. However, when the distance between
controllers generating the perturbations becomes large, control fails due to
increasing sensitivity of the system to noise and nonlinearities. We show that
this failure is due to the fact that the evolution operator for the controlled
system becomes increasingly nonnormal as the distance between controllers
grows. This nonnormality is the result of control and can arise even for
systems whose evolution operator is normal in the absence of control.

\end{abstract}

}

\newpage

Control of spatially extended systems has recently emerged as a problem of 
fundamental interest as well as significant technological importance. Numerous
investigations have shown the possibility of controlling a system by applying
feedback at every point in space. However, the practical implementation of
such control algorithms is often prohibitively expensive and sometimes
impossible. As a result, increasing attention is being devoted to localized
control, where feedback is applied only at a few spatial locations.  Previous
studies have shown that the minimal number of such control locations depends
both on the symmetry of the system \cite{symmetry} and the strength of noise
\cite{cml,egolf}.

The failure of localized control for large separation between controllers was
attributed to the phenomenon of nonnormality in the evolution operator
\cite{egolf}. For nonnormal systems, stability becomes a poor predictor of
short term dynamics \cite{farrell}. Strong nonnormality, which makes the system
extremely sensitive to noise, was previously identified as the mechanism that
provokes the transition to turbulence in uncontrolled systems, such as pipe or
channel flows \cite{trefethen,baggett}. The latter systems are nonnormal due to
a large mean flow, whereas generic extended systems have no mean flow and are
normal for typical boundary conditions. In these normal systems, nonnormality
arises as a result of control. Studies of nonnormality caused by localized
control \cite{egolf} have so far been limited to the interplay between
nonlinearity and the noise amplification due to nonnormality, rather than the
emergence of nonnormality itself. Our goals here are to investigate how
nonnormality arises in an originally normal system and study how nonnormality
leads to noise amplification, which determines the limits of localized control.

Let us consider the Ginzburg-Landau equation (GLE)
\begin{equation}
\label{eq_gl}
\dot{\phi}(x,t)=\phi(x,t)+\phi''(x,t)-\phi^3(x,t)
\end{equation}
as a prototypical system. Although simple enough to allow analytical treatment,
the GLE describes a very generic reaction-diffusion system. We therefore expect
the main results of the following analysis to apply to most extended dynamical
systems of this type.

The unbounded system (\ref{eq_gl}) possesses an unstable uniform steady state,
$\phi=0$. Our control objective is to make this state stable.
The symmetry of an unbounded system requires at least two
independent controllers in order to control a uniform target state
\cite{symmetry}, so we will break the symmetry by imposing certain boundary
conditions. Specifically, we will require that $\phi$ vanishes on one of the
boundaries, e.g., $\phi(0,t)=0$,  reducing the minimal number of controllers to
one. The only controller will be placed at the opposite boundary, $x=l$.
With this arrangement the length $l$ of the system plays the role of the
distance between multiple controllers in a system of larger size. We choose the
feedback law to be of the form
\begin{equation}
\label{bc_control}
\phi'(l,t)=\int_0^l K(y)\phi(y,t)dy,
\end{equation}
such that the control signal depends on the state $\phi$ of the system on the
whole domain (this condition can be easily relaxed). The feedback gain $K(y)$
describes how each point inside the domain contributes to the feedback.

The spectrum of the unperturbed linearized system is discrete with eigenvalues
and eigenfunctions given by
\begin{equation}
\label{eq_spectrum}
\lambda_n=1-q_n^2,\qquad f_n(x)=\sin(q_n x),
\end{equation}
where $q_n=(n-1/2)\pi/l$ and $n=1,2,\cdots$. These eigenfunctions are normal,
as expected, and form a convenient basis for the stability analysis of the
perturbed system. Dropping the cubic term in (\ref{eq_gl}) and projecting the
remainder onto the basis $\{f_n\}$ we obtain
\begin{equation}
\label{eq_gl_four}
\dot{\Phi}_n=\lambda_n\Phi_n-(-1)^n\sum_{m=1}^\infty K_m\Phi_m\equiv(M\Phi)_n,
\end{equation}
where $\Phi_n$ and $K_n$ are the Fourier coefficients of $\phi$ and $K$
\begin{eqnarray}
\Phi_n&=&\frac{2}{l}\int_0^l\phi(x)\sin(q_nx)dx,\nonumber\\
K_n&=&\frac{2}{l}\int_0^lK(x)\sin(q_nx)dx,
\end{eqnarray}
and
\begin{equation}
\label{eq_gl_mat}
M=\left(\matrix{
\lambda_1+K_1 & K_2 & K_3 & \cdots \cr
-K_1 & \lambda_2-K_2 & -K_3 & \cdots \cr
K_1 & K_2 & \lambda_3+K_3 & \cdots \cr
\vdots & \vdots & \vdots & \ddots}\right).
\end{equation}
The control problem for the PDE (\ref{eq_gl}) is thus reduced to finding an
infinite set of coefficients $K_m$, which will make the matrix $M$ stable.

It turns out that the structure of $M$ simplifies the problem remarkably. 
Suppose we take an $n\times n$ truncation of $M$ by discarding all rows and
columns except the first $n$.  Setting the eigenvalues of the truncated matrix,
$M_n$, to a sequence $\lambda'_1,\lambda'_2,\cdots,\lambda'_n$ is then
equivalent to solving a system of $n$ equations linear in $K_m$'s. In
particular, one can change the first $m=1,\cdots,s$  eigenvalues from
$\lambda_m$ to $\lambda'_m$ and leave the rest unchanged by setting
\begin{equation}
\label{eq_gain1}
K_m^{(n)}=-\frac{\prod_{p=1}^s(\lambda_m-\lambda'_p)}
{\prod_{p=1}^{m-1}(\lambda_p-\lambda_m)
\prod_{p=m+1}^{s}(\lambda_m-\lambda_p)},
\end{equation}
for $m\le s$ and zero otherwise. Since the right-hand-side does not depend on
the size of the truncated matrix $M_n$, the result also holds for the full
matrix $M$, so we can drop the index $n$. This is a very important result,
because it allows us to calculate Fourier coefficients of {\it any} stabilizing
feedback gain. It also allows us to determine how these coefficients scale with
the size of the system $l$. Substituting (\ref{eq_spectrum}) into
(\ref{eq_gain1}) and after some  algebra we obtain
\begin{equation}
\label{eq_gain2}
K_m=-\left(\frac{l}{\pi}\right)^{2(s-1)}
\frac{(2m-1)\prod_{p=1}^s(\lambda_m-\lambda'_p)}{(s+m-1)!(s-m)!}.
\end{equation}

\begin{figure}[t]
\centering
\mbox{\epsfig{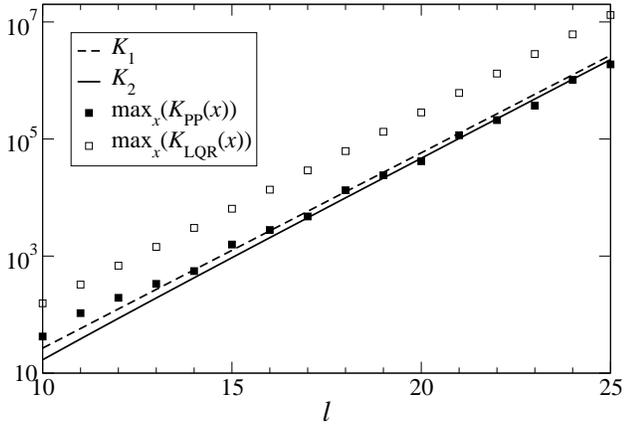}}
\vskip 3mm
\caption{Strength of feedback as a function of the size $l$ of
the system. The maximum of $K(x)$ in real space is compared with
the  Fourier coefficients $K_1$ and $K_2$ calculated using
(\ref{eq_gain3}).}
\label{fig_gain}
\end{figure}

To make the matrix $M$ stable we only need to change the
first $s$ positive eigenvalues, where $s$ is equal to the integer
part of $l/\pi+1/2$. In particular, in the limit
$\lambda'_1=\lambda'_2=\cdots=\lambda'_s=\Lambda$, where $\Lambda$
is some negative number, the product in the numerator of
(\ref{eq_gain2}) reduces to $(\lambda_m-\Lambda)^s$. Expressing
$s$ through $l$ we see that the coefficients $K_m$ grow
exponentially fast with $l$. The leading order behavior for large
$l$ is given by
\begin{equation}
\label{eq_gain3}
K_m\sim\exp\left[\frac{l}{\pi}\left\{2+\log(\lambda_m-\Lambda)
+O\left(\frac{m^2}{l^2}\right)\right\}\right].
\end{equation}
This exponential growth is clearly seen in Fig. \ref{fig_gain}. Both here and
throughout the paper we use and compare two different control methods. In
linear-quadratic (LQR) control \cite{dorato} the system (\ref{eq_gl_four}) is
truncated to 64 Fourier modes and a set of $K_m$ that minimizes a
quadratic form in $\Phi$ is sought numerically. In pole placement (PP) the
feedback gain is calculated directly from (\ref{eq_gain2}), where we change all
unstable eigenvalues to $\Lambda=-0.5$ and leave the stable ones unchanged. 
(The choice of $\Lambda$ is somewhat arbitrary and is chosen to roughly
correspond to the average of the first $s$ eigenvalues produced by LQR). 
Both control laws show the same scaling of $K$ with $l$. 

\begin{figure}[t]
\centering
\mbox{\epsfig{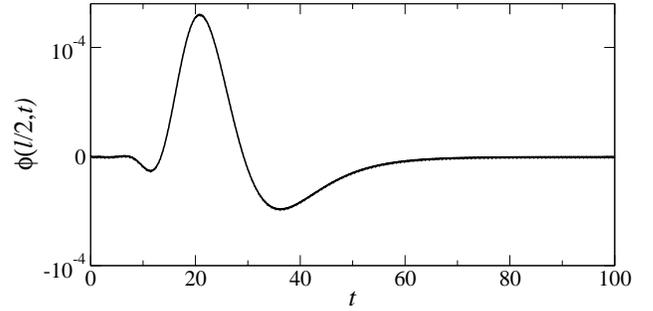}}
\vskip 3mm
\caption{The magnitude of the disturbance in the middle of a large domain,
$l=25$, under LQR control. Exponential asymptotic decay is preceded by a
transient, amplifying the initial disturbance (white noise with standard
deviation $\sigma=10^{-10}$) by six orders of magnitude.}
\label{fig_tran}
\end{figure}

The exponential growth of the control signal suggests transient behavior, a
sign of nonnormality. Indeed, a small initial disturbance inside the domain
will create a large control perturbation at the right boundary, $x=l$. If the
feedback is designed properly, this perturbation will eventually (after
propagating through the system) cancel the initial disturbance, thereby making
the system asymptotically stable. However, asymptotic decay will be preceded by
a transient whose magnitude grows with the feedback gain $K$. Numerical
simulation of the linearized GLE  does indeed show a large transient (see Fig.
\ref{fig_tran}).

The scaling of both the transient amplification and the control signal can also
be understood qualitatively. Since the propagation of perturbations is
diffusive, it will take a time $\tau$ roughly proportional to the size of the
system, $l$, for the control signal to travel from the right boundary to the
left one, suppressing the disturbances inside the domain. Since the system is
locally unstable, a disturbance near the left boundary will grow uncontrolled
during this time interval. The exponential growth of the disturbance will
result in its amplification by a factor $\exp(\lambda_1\tau)\sim\exp(\beta l)$.
To suppress the amplified disturbance, we need to apply a control perturbation
at least as large as the disturbance itself, which requires exponential (with
$l$) growth in the feedback gain.

Next, let us consider how nonnormality arises in our system. The evolution
matrix for the controlled dynamics can be
written in the form $M=A+BK^\dagger$, where $A$ is the diagonal (and hence
normal) matrix that describes the dynamics of the unperturbed system,
$A_{mm}=\lambda_m$, and $B$ and $K$ are vectors with elements $(-1)^{m+1}$ and
$K_m$, respectively. Clearly $BK^\dagger$  is not a normal matrix, so the sum
$A+BK^\dagger$ is not normal either. In fact, when all unstable eigenvalues are
chosen equal, the matrix $M$ is not even diagonalizable.  In that case, one can
convert $M$ into the Jordan normal form
\begin{equation}
\label{eq_jordan}
J(M)=S^{-1}MS=\left(\matrix{J_1 & 0 \cr 0 & J_2}\right)
\end{equation}
with $J_1$ being an $s\times s$ Jordan block with eigenvalue $\Lambda$,
$J_2$ a diagonal matrix with $\lambda_{s+1},\lambda_{s+2},\cdots$ on
the diagonal, and $S$ the respective coordinate transformation. This means
that the eigenvectors of $M$ corresponding to the Jordan block $J_1$ coincide,
$e_1=e_2=\cdots=e_s$. The solution of the linear system (\ref{eq_gl_four}) in
this case is constructed using the generalized eigenvectors $e'_p$, such that
$Me_p'=\Lambda e_p'+e_{p-1}'$ for $p=2,\cdots,s$ and $e'_1=e_1$. Specifically,
\begin{eqnarray}
\label{eq_gl_jord}
\Phi(t)&=&\sum_{p=1}^{s}\left[\sum_{m=0}^{s-p}c_{p+m}
\frac{t^m}{m!}\right]\exp(\Lambda t)e'_{p}\nonumber\\
&+&\sum_{p=s+1}^\infty c_p\exp(\lambda_pt)e_p,
\end{eqnarray}
where $c_1,c_2,\cdots$ are integration constants which have to be chosen to
satisfy the initial condition. 
The result for the individual Fourier coefficients can be written more
conveniently using the elements of the transformation matrix $S$:
\begin{eqnarray}
\label{eq_gl_comp}
\Phi_n(t)&=&\sum_{p=1}^{s}S_{n,p}\left[\sum_{m=0}^{s-p}c_{p+m}
\frac{t^m}{m!}\right]\exp(\Lambda t)\nonumber\\
&+&\sum_{p=s+1}^\infty S_{n,p}c_p\exp(\lambda_pt).
\end{eqnarray}
By looking at (\ref{eq_gl_jord}) or (\ref{eq_gl_comp}) one can clearly see that
the solution $\phi(x,t)$ grows as a polynomial of order $s-1$ before
exponential decay at a rate $\Lambda$ finally takes over.

\begin{figure}[t]
\centering
\mbox{\epsfig{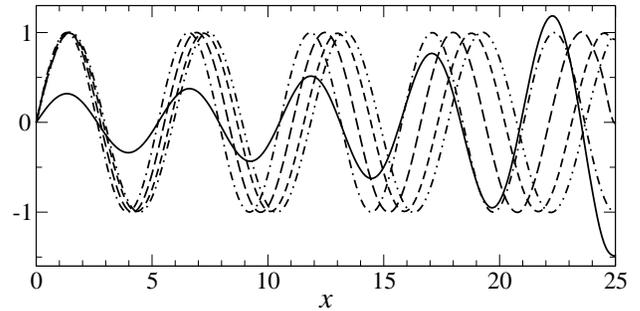}}
\vskip 3mm
\caption{First five eigenfunctions of the controlled system for $l=25$. The
evolution matrix $M$ was calculated using LQR and has all eigenvalues distinct.}
\label{fig_mode}
\end{figure}

In the above analysis we required the first $s$ eigenvalues to be equal.
What happens in the typical case when these eigenvalues are close, but not
equal? First of all, as (\ref{eq_gain2}) shows, the coefficients $K_m$ of
the feedback gain will still grow exponentially with the size of the system.
The matrix $M$ will remain strongly nonnormal, but will become diagonalizable.
Therefore, all eigenvectors of $M$ will be distinct, but the first $s$
will be closely aligned. This can be argued in the following way. Since
$K_m$ quickly grows with increasing $|\lambda'_m|$, while the size of the
largest possible perturbation is usually severely restricted by practical
limitations as well as nonlinearities, the new eigenvalues have to be chosen in
a small strip of negative values, say $(-\epsilon,0)$. Therefore, the increase
in the system size will force progressively more eigenvalues to lie in this
strip, making the difference between successive eigenvalues shrink at least as
fast as $\epsilon/l$. As we have seen previously, setting $s$ eigenvalues
equal produces an $s\times s$ Jordan block, which causes $s$ eigenvectors to
merge. Since $s$ is arbitrary, such merging will occur for any number of
eigenvectors corresponding to identical eigenvalues. As a result, pairs of
successive eigenvalues will continuously approach each other, aligning the
respective eigenvectors. (The continuity can be checked by a straightforward
application of perturbation theory.) Fig. \ref{fig_mode} shows that already for
$l=25$ the first five eigenfunctions are very closely aligned: all have a
nearly sinusoidal shape with either 8 or 9 nodes, i.e., almost coincide with
the first stable (unperturbed) eigenfunction $f_9(x)= \sin(8.5\pi x/l)$ which
has 8 nodes.

Finally, it is useful to derive the quantitative result for transient
amplification,
\begin{equation}
\label{eq_gamma}
\gamma\equiv\max_{t,\Phi(0)}\frac{||\Phi(t)||_2}{||\Phi(0)||_2}
=\max_t||\exp(Mt)||_2,
\end{equation}
because this is the ultimate measure that determines when modal analysis and
linear control break down. Let us again assume that all unstable eigenvalues
are made equal and analyze the structure of the solution (\ref{eq_gl_comp})
more carefully.  The transient occurs because each of the terms
$(t^m/m!)\exp(\Lambda t)$ first grows as $t^m$ and then decays as $\exp(\Lambda
t)$, reaching the maximal value at $t_m=-m/\Lambda$. This maximal value is
given by
\begin{equation}
\label{eq_tmax}
\frac{(t_m)^m}{m!}\exp(\Lambda t_m)\sim\frac{1}{\sqrt{2\pi m}\,|\Lambda|^m},
\end{equation}
so for small $|\Lambda|$ the term with $m=s-1$ dominates (\ref{eq_gl_comp}). A
good approximation for the transient amplification $\gamma$ can, therefore, be
obtained by picking the initial state $\Phi(0)$ that corresponds to setting
$c_m=\delta_{m,s}$ and calculating the maximum of the ratio $||\Phi(t)||_2/
||\Phi(0)||_2$:
\begin{equation}
\label{eq_gamma_0}
\gamma\sim\sqrt{\frac{\sum_{n=1}^s S_{n,1}^2}{\sum_{n=1}^s S_{n,s}^2}}
\max_{t}\frac{t^{s-1}e^{\Lambda t}}{(s-1)!},
\end{equation}
where we only keep terms with $p=1$.
It should be noted that the chosen initial state corresponds to a ``near
optimal'' disturbance, rather than the ``optimal'' disturbance producing the
largest transient amplification \cite{farrell}.

If the matrix S is normalized such that $S_{1,s}=1$, it can be shown that,
for arbitrary $n$, $S_{n,s}=\delta_{n,1}$ and
\begin{equation}
\label{eq_s11a}
S_{n,1}=(-1)^{s+n}\frac{(\lambda_1-\Lambda)^s}{\lambda_n-\Lambda}
\frac{\prod_{k=2}^s(\lambda_k-\Lambda)}{\prod_{k=2}^s(\lambda_1-\lambda_k)}.
\end{equation}
Since all $S_{n,1}$ scale in the same way, (\ref{eq_gamma_0}) gives
\begin{equation}
\label{eq_gamma_1}
\gamma\sim\frac{|S_{1,1}|}{|\Lambda|^{s-1}}.
\end{equation}
In order to perform the calculation of the leading order behavior of $S_{1,1}$
we assume that $\Lambda=1-[(p-1/2)\pi/l]^2$, where $p$ is some integer, in
which case (\ref {eq_s11a}) gives
\begin{equation}
\label{eq_s11b}
|S_{1,1}|=\frac{(\lambda_1-\Lambda)^{s-1}(p+s-1)!}{(p-s-1)!p(p-1)s!(s-1)!}.
\end{equation}
We can now re-express $p$ in terms of $\Lambda$ thus extrapolating between the
known expressions for $\Lambda$'s corresponding to integer $p$'s. At the
leading order we again obtain exponential growth with the system size:
\begin{equation}
\label{eq_gamma_2}
\gamma\sim\exp\left[\frac{l}{\pi}\left\{\log\frac{\Lambda-1}{\Lambda}+
\log\frac{(\alpha+1)^{\alpha+1}}{(\alpha-1)^{\alpha-1}}\right\}\right],
\end{equation}
where $\alpha\equiv\sqrt{1-\Lambda}$.

\begin{figure}[t]
\centering
\mbox{\epsfig{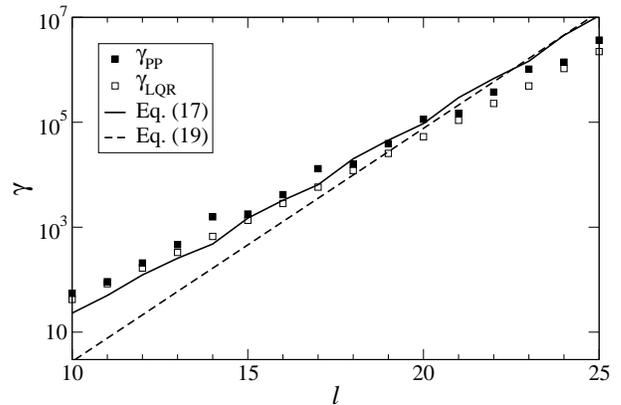}}
\vskip 3mm
\caption{Transient amplification $\gamma$ as a function of $l$. The squares 
show the values calculated numerically using (\ref{eq_gamma}).}
\label{fig_ampl}
\end{figure}

As Fig. \ref{fig_ampl} shows, the numerically calculated transient
amplification factor does indeed grow exponentially fast and is rather
insensitive to the way the feedback gain is calculated. The slope is seen to be
slightly different from the one predicted by (\ref{eq_gamma_2}). This is to be
expected.  First, we only consider the terms in (\ref{eq_gl_comp}) for which
$m=s-p$. While that gives a correct leading order result for small $|\Lambda|$,
for $\Lambda=-0.5$ the contribution from $m=s-p-1$ is about half that from
$m=s-p$ and more terms might need to be considered. Second, (\ref{eq_gamma_2})
is the asymptotic result for (\ref{eq_gamma_0}) and is valid only in the limit
of large $l$ (here for $l\gtrsim 20$). On the other hand, the numerical
accuracy in calculating the matrix norm decreases rapidly with $l$. This is in
fact a numerical fingerprint of nonnormality. The results found using standard
numerical routines are getting rather inaccurate for strongly nonnormal
matrices, e.g., for large $l$ (in our case also for $l\gtrsim 20$).

The large transient amplification makes the system extremely susceptible to
noise, as noise is amplified by feedback before being suppressed. In order for
linear control to work, the magnitude of the nonlinear terms has to be smaller
than the magnitude of the linear terms. Comparison of their relative magnitude
can be used to estimate when noise will start to interfere with control. For
instance, for cubic nonlinearity in the GLE, using the argument of Egolf and
Socolar \cite{egolf} we obtain that the largest magnitude of noise, $\sigma$,
tolerated by linear control should scale like $\gamma^{-3/2}$. However, our
numerical calculations produce different scaling. This disagreement is
currently under investigation.

To conclude, we have shown that the application of spatially localized control
to spatially extended systems renders the system nonnormal. The degree of
nonnormality quickly increases with the size of the system due to the close
alignment of a progressively larger number of originally unstable
eigenfunctions. Increasing nonnormality leads to transient amplification which
quickly grows with the size of the system, thus imposing strict limitations on
the density of controllers required to control a system of a given size in the
presence of noise or truncation errors.

\end{document}